\documentclass[a4paper,12pt]{article}
\usepackage{lastpage}
\usepackage{graphicx}
\pdfoutput=1
\usepackage{float}
\usepackage{xcolor}
\usepackage{multirow}
\usepackage{diagbox}
\usepackage[toc,page,title,titletoc,header]{appendix}
\usepackage{listings}
\usepackage{amsmath}
\usepackage{pxfonts}
\usepackage{mathrsfs}
\usepackage[numbers,sort&compress]{natbib}

\usepackage{theorem}
\usepackage[scale=0.75]{geometry}

{
  
\usepackage[scale=0.75]{geometry}
\usepackage{bbding}
\usepackage{url}
\usepackage{booktabs,longtable}
\usepackage{mdwlist}
\usepackage{subfigure}
\usepackage{pifont}
\usepackage[labelsep=space]{caption}
\usepackage{multirow,paralist}
\usepackage[title,titletoc]{appendix}
\usepackage{amsfonts}
\usepackage{color}
\usepackage[pdftex,colorlinks=true]{hyperref}

\usepackage{flafter}
\usepackage{pifont}
\usepackage{picinpar}
\usepackage{indentfirst}    
\usepackage{longtable}      
\usepackage{subfigure}

\usepackage{mathptmx}  
\usepackage{mathpazo}  
\usepackage{helvet}    

\usepackage{colortbl}
\definecolor{grey}{rgb}{0.91,0.91,0.91}

\definecolor{grey}{rgb}{0.3,0.3,0.3}
\definecolor{darkgreen}{rgb}{0,0.3,0}
\definecolor{darkblue}{rgb}{0,0,0.3}
\definecolor{grey}{rgb}{0.95,0.95,0.95}
\definecolor{darkgreen}{rgb}{0.0,0.5078,0.0}

\lstset{language=Matlab}
\lstset{xleftmargin=1em,xrightmargin=1em}
\lstset{frame=}
\lstset{
commentstyle=\color{darkgreen},keywordstyle=\color{blue},
breaklines=true,columns=flexible,mathescape=fause }
\lstset{framexleftmargin=0.5em,framexrightmargin=1em,framextopmargin=0em,basicstyle=\footnotesize,framexbottommargin=1em}
\lstloadlanguages{C,C++,Java,Matlab,Mathematica}

\usepackage{titletoc}
\titlecontents{section}[0mm]
{\vspace{.2\baselineskip}\bfseries}
{\normalsize{\thecontentslabel}~\hspace{.5em}} {}
{\dotfill\contentspage[{\makebox[0pt][r]{\thecontentspage}}]}
[\vspace{.1\baselineskip}]
\dottedcontents{section}[1.16cm]{\large}{2.0em}{4pt}
\dottedcontents{subsection}[2.00cm]{\normalsize}{2.0em}{4pt}

\usepackage{CJK}
\begin{document}

\date{}
\title{Research on two-dimensional traffic flow model based on psychological field theory}
\maketitle

\begin{center} {Wenhao Li$^1$,\ Yu Nie$^2$,\ Zhongyao Yang$^2$,\ Shaochun Zheng$^1$\ and~~Dehui Wang$^1$}
\end{center} \begin{center} {\emph{ $^{1,2}$College of Mathematics and College of Computer Science and Technology, Jilin University, Changchun, China}} \end{center}
\newcommand{\upcite}[1]{\textsuperscript{\textsuperscript{\cite{#1}}}}
\bibliographystyle{plain}


{\textbf{Abstract}}:\ In this paper, the influence of fan-shaped buffer zone on the performance of the toll plaza is researched. A two-dimensional traffic flow model and a comprehensive evaluation model based on mechanical model and psychological field are established. The traffic flow model is simulated by creating coordinate system.

We first establish queue theory model to analyze vehicles when entering toll plaza. Then, a two-dimensional steadily car-following model is established based on psychological field for the analysis of vehicles when leaving toll plaza. According to psychological field theory, we analyze the force condition of each vehicle. The force of each vehicle is contributed by the vehicles in its observation area and obstacles. By projecting these vehicles and obstacles via the equipotential line in the psychological field, the influence on the value and direction acceleration
of following vehicles is obtained. Consequently, the changes of each vehicle¡¯s speed and position are obtained as well. Next, we establish simulation based on the states of vehicles and make the rules of vehicle state-changing.

By simulating the system, we obtain the throughput of the toll plaza¡¯s input and output. Then we obtained the bearing pressure on the road by the max throughput and the demand of the roads. Using the number of cars in per unit area as the safety factor. Then a comprehensive evaluation model is established based on bearing pressure on the road, cost and safety factor.

In simulation study, we obtain construction scheme of the toll plaza by substituting different values of parameters, such as max vehicle flow, lanes, toll booths and the length of the buffer area.Then we take automated vehicles and the different service efficiency of toll booth into consideration. After all, the simulation results under different conditions are used to substituted in the evaluation model. Finally, we offer a better construction scheme of toll plaza in the New Jersey according to various evaluation results.

\newpage


\section{Introduction}
With the rapid expansion of road transportation, more toll highways come into people's daily life\cite{Campbell}. There are mainly two kinds of toll stations known as ramp tolls and barrier tolls\cite{Wilson}. The barrier tolls are widely used and studied now. There are series of problems caused by this kind of toll points, such as high cost and long rehicular queue\cite{Rillings}.
A study shows that 36\% of traffic congestion on toll ways are caused by stopping to pay tolls\cite{Rong}. In this study, we shall address:

\subsection{Literature Review}
The existing studies on traffic problems can be divided into macroscopic and microscopic studies\cite{Prigogine}.

The macro-traffic model treats traffic as a fluid flow. In 1955, the British scholar Lighthill\upcite{Lighthill} and Whitham\upcite{Whitham} established a one-dimensional continuity equation of traffic flow. The macro-traffic model not only describes the distribution of traffic flow along the road space, but also reflects the law of time-varying traffic flow. It can accurately describe the real behavior of traffic flow, which has been widely used in actual traffic control.

Microscope models are discrete models of traffic flow which study the behavior of individual vehicles and drivers and use the distance between vehicles and the speed of vehicles to describe the traffic flow. For example, car-following model and Cellular Automaton model. When the traffic is heavy and the inter-vehicle spacing is very small, the speed of each car is affected by the speed of the car ahead. The driver can only control the speed of his car at the speed of the front car, which is called "steadily car-following". Steadily car-following theory which is an important theoretical base for microscopic traffic flow simulation\cite{YAO}. Since the 1950s,  based on the understanding of driver's behavior from different perspectives, researchers derived a series of steadily car-following models such as Stimulus-Response Model, Safety Distance Model, Psycho-Physical Model and Artificial Intelligence based Model\cite{Garber}. In the early 1990s, researchers began to use cellular automaton as a model of traffic flow in microscopic studies. The steadily car-following model and the cellular automaton model use the numerical simulation method to simulate the motion behavior of the vehicle from microscopic view.
After study, we find shunting of vehicles in the input system was much less important on the problem than the confluence of the vehicles in the output system\cite{Nagel K}.

%



\section{General Assumptions}

\begin{asparaenum}[(1)]
\item The same kind of vehicle behavior is consistent.
\item We use average service time of vehicles in the same toll station.
\end{asparaenum}



\section{Pull in System of Toll Booth}

The vehicles drive from one end of the highway into the toll station. Assuming that, in the unit time, the number of cars be ready to drive from the end of the highway to the toll station obeys Poisson distribution with parameter $\lambda$, where $\lambda$ represents the expected arrived number of cars arrived in one unit. The probabilities of i times arrived in one unit time is


\begin{eqnarray}\label{1}
P(X=i)=\frac{e^{-\lambda}\lambda^i}{i!}
\end{eqnarray}

\begin{figure}[!htbp]
\small
\centering
\includegraphics[width=12cm]{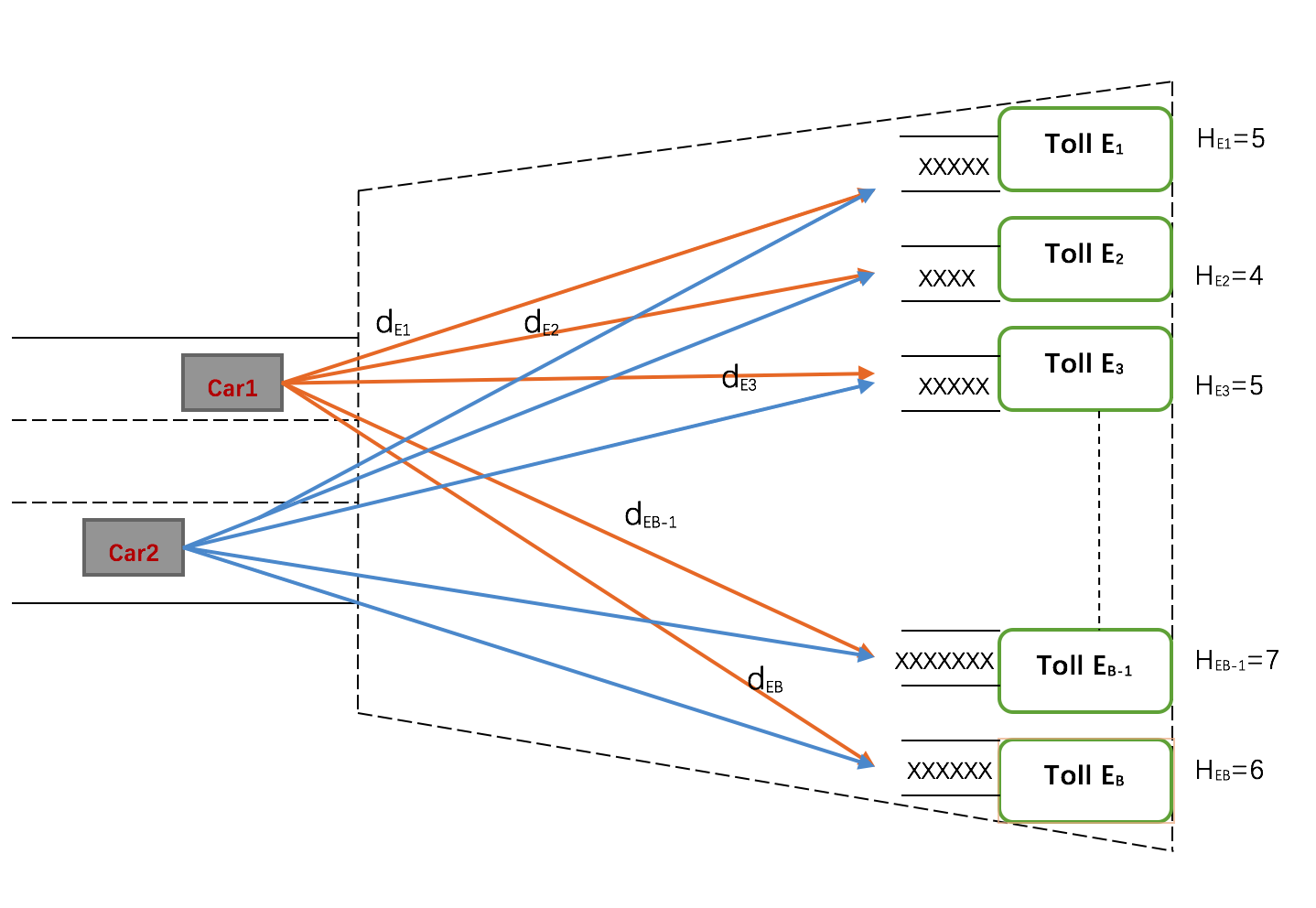}
\caption{:Input system of toll booth} \label{shuru}
\end{figure}

Before choosing a toll gate, each driver need to go through these steps:(see figure \ref{shuru})

\par

For each vehicle, his choice process of toll gate can be divided into the following three steps:


Step 1: Determine the current number of queuing cars from the toll station\ $E_1$\ to\ $E_n$, then change and choose the shortest queue. The number of cars is $H_{Ei}$.


Step 2: If there are more than two waiting line are the same length,  the straight distance between the vehicle and the toll station will determine the choice. The straight distance is $d_{Ei}$.

%

Step 3: If two toll stations have the same $d_{Ei}$, the driver shall choose the right toll station.


According to Edie's\upcite{Edie}, for each vehicle arrived at the toll station, the average service time $t_1$ is equal to 12 seconds. The average ETC service time is 2 seconds which is proved by Orlando-orange National Expressway web site. And Houston's report\cite{Houston} shows that, the average service time of self-service toll station is 7 seconds. When a vehicle drives out of the toll booth, we analysis the condition after toll. When using discrete time traffic flow model, we further assume each parking space as a grid. Then we can use the matrix$(\delta_{ij})_{m\times L}$ to represent the occupancy of parking spaces behind all the toll stations,
\begin{eqnarray}\label{sudubianhua}     
\delta_{ij}=\left\{                        
\begin{array}{lll}       
1,\ if\ the\ space\ is\ occupied \\  
0,  otherwise  . \\

\end{array}              
\right.                       
\end{eqnarray}
 Then we will get a matrix of m columns,  the i-line of the matrix represents the occupancy of parking spaces in the rear of toll booth(starting position of fan-shaped buffer)in the i-time.


\section{Basic Traffic Models Based on Psychological Field }

From the driver's psychology£¬we analyze the relationship of car and car, car and highway boundaries, as well as car and fan-shaped buffer boundary.
We first consider the driver's own situation, the driver's planning route will attract drivers, guiding the driver road. For example, drivers who want to exit the current road as soon as possible will be attracted by the road exit. We can assume that the vehicle is subject to traction from the direction of the exit.  Similarly, in the vehicle and the front of the vehicle or obstacle, there exists an force due to the driver's psychological effect. An obstacle may pose a threat to the driver's safety. The threat generated by the obstacle is abstracted as creating a repulsive force field around it, which makes repulsive interaction to the driver's vehicle. Therefore, based on the psychological field, we analyze the one-dimensional traffic flow model. Namely, we establish safety distance, vehicle speed model and collision avoidance model based on the driver's psychological effect.


We calculate the safety distance based on the psycho-field car following model. The concept of car-following was first proposed by Reuschel\upcite{Tomer} and pipes\upcite{Mark} in the 1950s. Kometani and Sasaki\upcite{E Kometani}  put forward the concept of safety distance car-following model: the vehicle needs to keep a certain distance, that is, the distance from the head of front vehicle to the trail of the rear vehicle, to avoid collision. In this paper, we use the safety distance under the NETSIM model\cite{Ajay K}: the vehicle has enough time to react and slow down or even stop in order to avoid collisions. The safety gap model is also known as the collision avoidance model. When the car-following speed is higher than the preceding vehicle, the distance between the two vehicles is reduced until reaching the driver's deceleration of critical threshold (safety distance). Based on the psychological point of view, in this time, because of the psychological role of the driver, the front car will make an acting force to rear cars, the following cars began to slow down in an attempt to keep pace with the speed of front cars. But the driver can not guarantee the accuracy of the operation, that is, the distance of the two car heads will be reduced to a certain extent, then the interval will increase until the driver reaches the acceleration threshold. At this point the driver will accelerate again and repeat the process. So the safety distance is also a driver's psychological safety distance.


Psychological safety distance can be divided into two categories, the safety distance between vehicles (dynamic safety distance) and the safety distance between vehicles and obstacles (obstructions are static).

\begin{figure}[!htbp]
\small
\centering
\includegraphics[width=8cm]{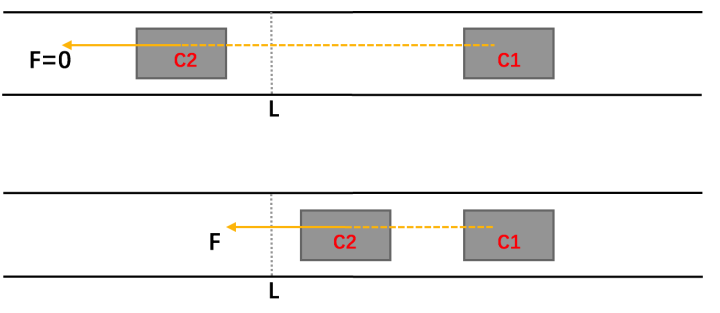}
\caption{:safety distance based on psychological field } \label{zhixiansafety}
\end{figure}
\par
The two parts in the figure\ref{zhixiansafety} are states of different time in the same road. We see two vehicles as\ $C_1$\ and\ $C_2$, $C_1$ for the preceding car. When $C_2$ is on the left side of the dashed line L, the psychology of the driver in $C_2$ is influenced by $C_1$ which is zero. When $C_2$ travels to the right of L, the driver's psychology in $C_2$ is affected by $C_1$. It is assumed that the distance from L to $C_1$ is a safe distance. When distance from $C_1$ to $C_2$ is greater than the safe distance, $C_2$ driver's psychology will not be affected by the force from $C_1$. When distance from $C_1$ to $C_2$ is less than the safe distance, $C_2$ driver's psychology will be affected by the force from $C_1$. It can be considered that there is a region around $C_1$, the vehicle driver will be affected by the force from $C_1$ if it in the region. This region can be compared to a force field.


First, we calculate the dynamic safety distance between vehicles. Note the two vehicles as\ $C_1$\ and\ $C_2$, $C_1$for the first car, assuming the speed of\ $C_1$\ and\ $C_2$, respectively, are the current vehicle speed $v_1$and\ $v_2$. When encountering the emergency situation, $C_1$ will decelerate with acceleration $a_1$, then $C_1$\ will stop in the $v_1/a_1$  time, at this time\ $C_1$\ covering distance $S_1=v_1^2/2a_1.$


Assuming that the reaction time of vehicle $C_2$ is t, the average reaction time\ $t=0.73$, which is studied by Tao\upcite{Tao}. That is, after t time to start braking and deceleration with the maximum acceleration\ $a_1$, it will stop in\ $t+ v_2/a_1$ time. The travel distance is $S_2= v_2 \cdot t + v_2^2/2 a_1.$


The distance between the two cars is reduced by\ $S_2 -S_1$, plus the vehicle length L. This distance is the minimum distance between vehicles to prevent collision, which can be defined as the safety distance G.


\begin{eqnarray}\label{1}
G=L+v_1\cdot t+\frac{v_2^2-v_1^2}{2 a_1}.
\end{eqnarray}

Then, we calculate the safety distance between the vehicle and the obstacle (the obstacle is static). At this time the car can decelerate with acceleration \ $a_2$, and the reaction time of the vehicle is expected reaction time, denoted by \ $T$, $T$\ is expected to be\ $0.54s$\cite{Tao}. We can also see this as the velocity of the preceding vehicle is 0, namely $v_1=0$. Obstacle is the vehicle that is static. So the safety distance G' is given as:


\begin{eqnarray}\label{1}
G'=v_1\cdot T+\frac{v_2^2}{2 a_2}.
\end{eqnarray}

We establish a vehicle acceleration model, the vehicle will adjust their driving mode based on the preceding cars' situation or obstacles . The vehicle will choose to continue to maintain the current speed, accelerate or decelerate.


According to road traffic knowledge, we know that the maximum speed of vehicles in the toll booth fan-shaped buffer zoneis\ $v_{max}$\cite{Traffic safety psychology}. When the vehicle is traveling at maximum speed and there is no vehicle or obstacle within a safety distance, the vehicle is running at the current speed.


If the distance between the following vehicle and the preceding vehicle or obstacle is greater than the safety distance, that is, the preceding vehicle or obstacle does not have a psychological effect on the vehicle driver. And if the vehicle does not reach the maximum velocity $v_{max}$, it will accelerate with acceleration\ $a_3$\ to achieve maximum speed or make the distance less than the safety distance from the preceding vehicle or obstacle.


When the distance of the following vehicle from the preceding vehicle or obstacle is less than the safety distance, the following vehicle \ $C_2$\ has the risk of collision with the preceding vehicle. Under the role of psychological field, the preceding car will give the following car a reverse force. That is, the following car needs to slow down so that it will keep the distance from the preceding car greater than the safety distance. The acceleration required for the following vehicle\ $C_2$\ is denoted by\ $a_4$. Assume that the speed and position of the preceding vehicle\ $C_1$\ at the present time are\ $v_1$\ and\ $x_1$, respectively. The speed and position of the following vehicle are\ $v_2$\ and\ $x_2$, respectively. After\ $\Delta t$ , new location and speed of\ $C_1$\ and\ $C_2$\ are respectively\ $v'_1$\ \ $x'_1$\ and\ $v'_2$\ \ $x'_2$. At the same time, we can get a new psychological safety distance\ $G'=L+tv'_1+\frac{{v'}_1^{2}-{v'}_2^{2}}{2 a_1}$. And the distance difference between\ $C_1$\ and\ $C_2$\ after\ $\Delta t$ is \ $x'_1-x'_2\ge G'$. Therefore, we can draw the conclusion that the minimum acceleration of the following car\ $C_2$\ in the psychological field car-following model is:

\begin{eqnarray}\label{1}
a_4&=&\frac{1}{\Delta t}(\frac{\Delta t a_1}{2}+v_1+ta_1-\frac{1}{2}[(\Delta ta_1)^2+4\Delta t a_1^2 t+(2ta_1)^2-4v_1 \Delta t a_1 \nonumber \\
&+& 8(x_2-x_1)a_1 +8 v_2 \Delta t a_1 -8La_1+4v_2^2]^{\frac{1}{2}}).
\end{eqnarray}

When there is an obstacle in front, let \ $v_2=0$, $L=0$, the reaction time is T, the acceleration is $a_2$. Hence, we have

\begin{eqnarray}\label{1}
a_5=\frac{1}{\Delta t}(\frac{\Delta t a_2}{2}+v_1+Ta_1-\frac{1}{2}[(\Delta ta_2)^2+4\Delta t a_2^2 T+(2 T a_2)^2-4v_1 \Delta t a_2 +8(x_2-x_1)a_2 ]^{\frac{1}{2}}).
\end{eqnarray}

\section{The Car-following Model of Two-dimension Based on Psychological Field}

\subsection{The Equipotential Line of Psychological Field}

\begin{figure}[ht]
 \begin{minipage}[ht]{0.5\linewidth}
 \centering
 \includegraphics[width=0.9\textwidth]{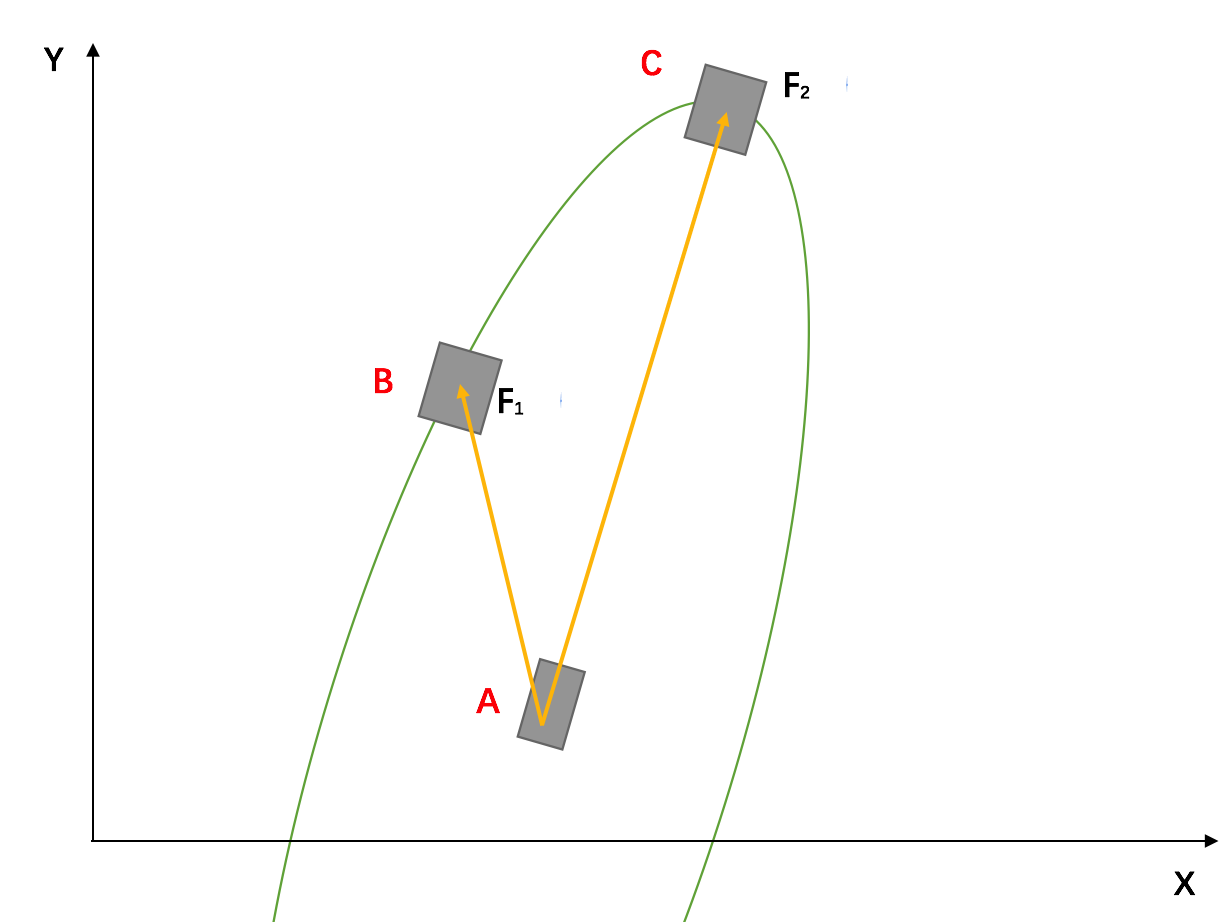}
 \caption{:psychological field of carA, $F_1=F2$} \label{xinlichang}
 \end{minipage}
 \begin{minipage}[ht]{0.5\linewidth}
 \centering
 \includegraphics[width=0.8\textwidth]{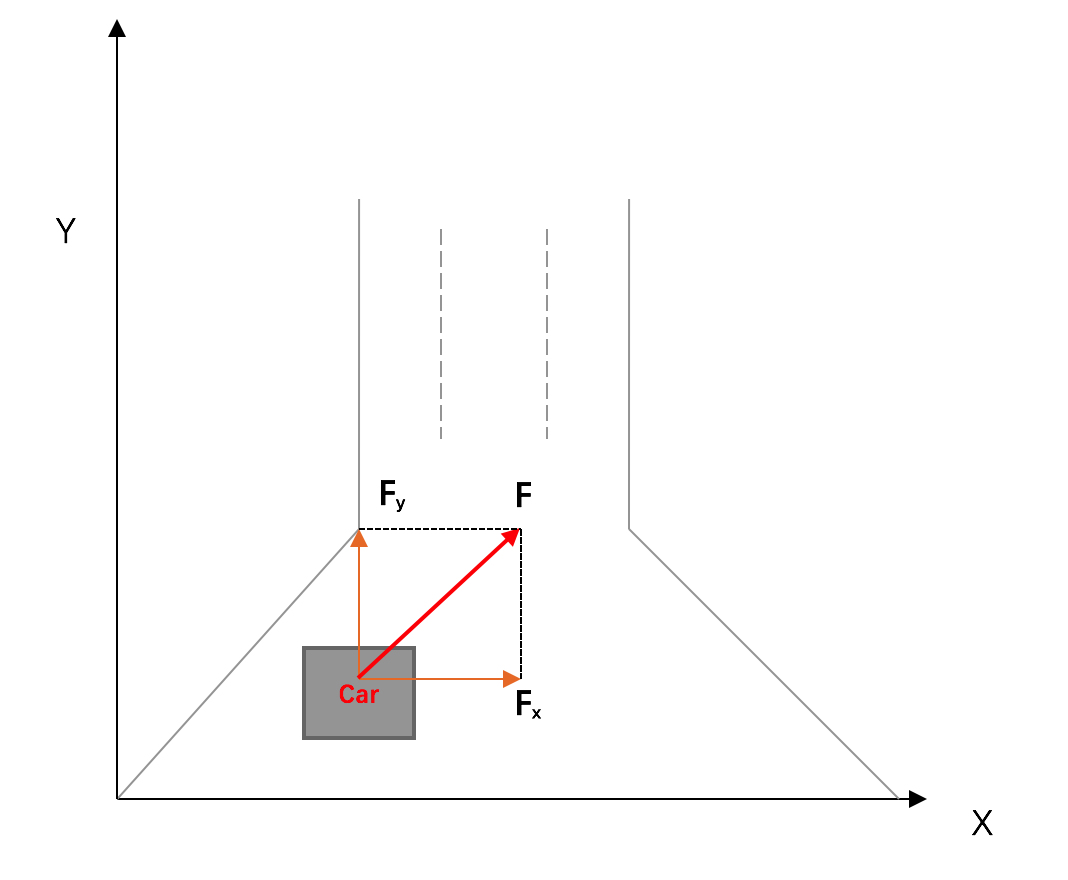}
 \caption{:force analysis of car and destination} \label{zuoyonglizishen}
\end{minipage}
\end{figure}

In the plane region, the psychological field is a mathematical abstraction of psychological pressure which drivers generate in the process of driving with being affected by the surrounding environment. This psychological pressure is related to the distance between the driver and the surrounding objects.
From the above discussion we know that in a one-dimensional line, the field strength size is inversely proportional to the length. On the two-dimensional plane, however, the preceding vehicle has an offset distance from the car-following. Offset distance is greater, the the force of front vehicle on the car is less. It can be said that the influence of the leading vehicle on the car-following car is related to the distance between the vehicle and the vehicle in the car-following direction. As the speed is fast, drivers will weaken their concern on both sides. So the area that will affect the driver tends to be flat (half elliptical area). So we can introduce the concept of equipotential line(see figure\ref{xinlichang}). The shape of the equipotential line is half elliptical, and all obstructions on this line have the same effect on the car. In other words, it is based on the same force under the psychological field. Further, we can project the forces between the vehicles on the major axis of the ellipse according to the equipotential lines. That is to project the preceding car to the car-following speed direction(see figure\ref{xinlichang}). For example, project the car B to the car C. Because it is the projection along the equipotential line, so keep the same force on the car-following force , namely $F_1 = F_2$. Based on this, we can project all the vehicles having an impact on the follow-car vehicles to the speed of car-following direction (ray AC). Line AC is the semi-major axis of the ellipse, denoted by a, remember the corresponding semi-short axis is b. The angle between AB and AC is\ $\theta$, $\theta \in [0,\pi/2]\cup[3\pi/2,2\pi]$. The distance between AB is denoted by $\rho$. Hence, we have


\begin{equation}\label{tuoyuantouying1}
\rho = a-(1-\frac{b}{a})a|sin\theta|.
\end{equation}

Further, let $\alpha=\frac{b}{a}$  be the shape parameter of the equipotential line. $\alpha$ is related to the speed of the vehicle. The faster the vehicle travels, the flatter the shape of the equipotential line is, that is, the smaller the $\alpha$ is. The slower the vehicle travels, the smoother the shape of the equipotential line is, the greater the $\alpha$ is. When we give a certain car-following speed, $\alpha$ is a fixed value. We can compute the projection position of the vehicle, namely, the semi-major axis a of the elliptical projection is a.


\begin{equation}\label{tuoyuantouying2}
a = \frac{\rho}{1-(1-\alpha)|sin \theta |}.
\end{equation}
$(u_A,v_A)$ is the speed of car A, and the coordinate of vehicle A and B are $(x_A,y_A)$ and $(x_B,y_B)$, so we can get the distance $L_s$ from car B to AC, that is, the offset distance.

\begin{equation}\label{tuoyuantouying3}
L_s=\frac{v_A x_B-u_A y_B -v_A x_A+u_A y_A}{(u_A^2+v_A^2)^{\frac{1}{2}}}.
\end{equation}
Further£¬we can obtain $sin \theta=L_s/\rho$.

\begin{figure}[ht]
 \begin{minipage}[ht]{0.5\linewidth}
 \centering
 \includegraphics[width=1\textwidth]{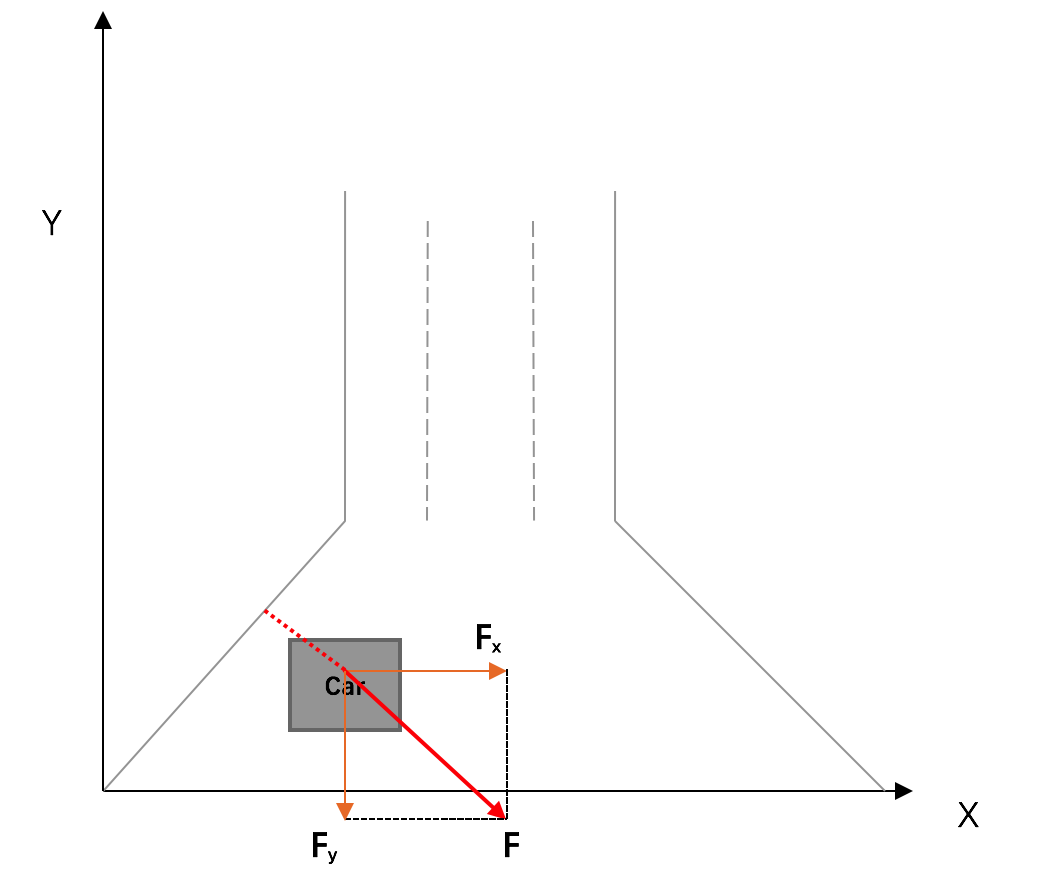}
\caption{:force analysis of car and  boundary} \label{zuoyongliqiang}
 \end{minipage}
 \begin{minipage}[ht]{0.5\linewidth}
 \centering
 \includegraphics[width=1.1\textwidth]{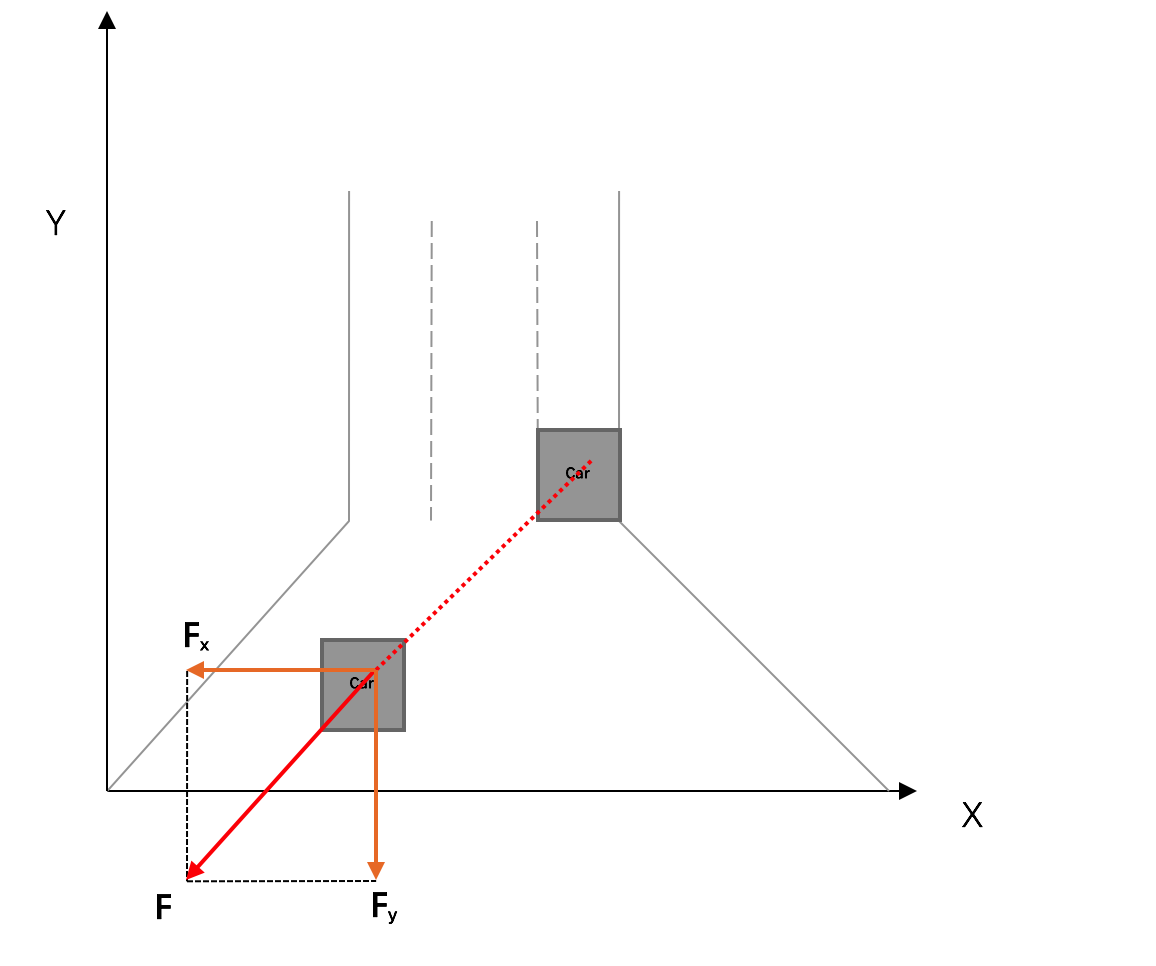}
\caption{:force analysis of cars} \label{zuoyongliche}
\end{minipage}
\end{figure}

In the psychological field, we carry on the force analysis to the vehicle. The driver would like to exit the fan-shaped buffer area as soon as possible, so car itself has a traction force F(see figure\ref{zuoyonglizishen}). F is decomposed into $F_x$ and $F_y$ by orthogonal decomposition. Car is subjected to a repulsive force F' from the edge of the road(see figure\ref{zuoyongliqiang}), and F' is decomposed into $F'_x $ and $F'_y$ by orthogonal decomposition.
Car is also subject to a repulsion from the front of the vehicle(see figure\ref{zuoyongliche}), the same orthogonally decomposed to be $F''_x$ and $F''_y$.
The six forces in the x and y directions of the car are vector-added to obtain the component forces $f_x$ and $f_y$ of the resultant force of the three forces in the x and y directions.

Based on the psychological field model, a safety zone can be defined. We can use the equipotential lines to define the safety zone. Suppose that the obstacle on the equipotential line has zero effect on the car-following, and the effect of the obstacle outside the equipotential line is zero. So drivers only need to consider the safety zone of the obstacles. The projection of the safety zone in the velocity direction is the safety distance. In order to avoid the collision, the following car needs to make appropriate braking when an obstacle enters into the safe area. This means that the car will be affected by the opposite direction of force.


\subsection{Traffic Rules}

We develop the rules of the vehicle in the psychological field. Under the influence of the psychological field, car-following vehicles are subject to the forces of the preceding vehicle and the road boundary. This is a repulsive or attractive force. Road borders can be viewed as stationary vehicles. So the role of the vehicle and the wall can also be seen as the role of vehicles and vehicles. Assuming that the current force of the vehicle is $f$, the decomposition in the x and y directions is $f_x$ and $f_y$.
Using $f_x$ and $f_y$ , we can get car's $a_x$ and $a_y$ in the X and Y direction of acceleration, in the next time. Using the components of the current velocity in X and Y directions, $a_x$ and $a_y$, we can get the velocity u, v, and the position x, y of the next time car.
Assuming that the coordinate of car is (x, y), the current state of car can be represented by a four-tuple $(x,\ y,\ u,\ v)$ .


Let us first consider the changes in the speed of the vehicle in the psychological field. Suppose that the current speed of car-following is $(u_n(t),v_n(t))$, where $u_n(t)$ is the horizontal velocity and $v_n(t)$ is the vertical velocity. And the current acceleration $a_n(t)=(a_{nx}(t),a_{ny}(t))$. $(x_n(t),y_n(t))$ is the current coordinate of car-following, where, $x_n(t)$ is the abscissa and $y_n(t)$ is the ordinate.
The current speed of the preceding vehicle is $(u_{n-1}(t), v_{n-1}(t))$. The current coordinate of preceding vehicle is $(x_{n-1}(t),y_{n-1}(t))$, and $y_{n-1}(t)$ is the ordinate , $x_{n-1}(t)$  is the abscissa. The leading vehicle produces a force on the car-following vehicle and further produces acceleration. That is car-car acceleration in $t+T$ time, recorded as $a_n(t+T)$. Because the greater the speed difference is, the more the impact of leading vehicles will act on the car-following. The greater the force is, that is, the greater the acceleration will be. It can be summarized that the current vehicle speed is much greater than the rear, the car will speed up, and the acceleration increases. When the preceding vehicle speed is far less than car-following, the car-following will slow down and the deceleration will increase. On the other hand, the greater the distance between the leading vehicle and the following car is, the smaller the force of the leading vehicle on the car-following car is,that is, the smaller the acceleration will be. When the car-following speed increases in the $t+T$ time, the process of acceleration will become faster, that is, the acceleration will be greater. Thus, we have

\begin{equation}\label{an}
a_{n}(t+T)=\lambda v_{n}^m (t+T)\frac{\Delta v(t)}{\Delta x^l (t)},
\end{equation}
where $\lambda, m\ and\ l$ are constants, $\Delta v(t)$ is the velocity difference between the nth vehicle and the $(n-1)$th vehicle at time t. $\Delta x^l (t)$ is the distance between the n-th car and the $(n-1)$th car at time t.  $v_{n} (t+T)$ is the speed at which $t+T$ is following the car.


We present (\ref{an}) with the current velocity and acceleration of a car-following vehicle, and further introduce the psychological and equipotential lines. We have

\begin{equation}\label{anxinlichang}
a_{n}(t+T)=\lambda[ \frac{v_{n} (t)+a_n(t)T}{\Delta x (t)}(1-(1-\alpha)\frac{L_s}{\Delta x (t)})]^m \frac{\Delta v(t)}{\Delta x^{l-m} (t)},
\end{equation}
where $L_s$ is the offset distance, and $\alpha$ is the ellipse shape parameter.


In this way, we extend the one-dimensional car-following model to a two-dimensional car-following model. And we consider the effect of the leading vehicle on the car-following force, that is, the influence of acceleration. We can use the projection method on an ellipse (based on equipotential lines) to analyze the magnitude of the forces acting on the car-following model. By (\ref{tuoyuantouying2}), at this point the size of all the leading vehicles have been projected to the car in the forward direction(see figure\ref{xinlichang}), that is, the new offset distance$L_s=0$. Thus£¬(\ref{anxinlichang}) simplifies to

\begin{equation}\label{anxinlichangjianhua}
a_{n}(t+T)=\lambda[ \frac{v_{n} (t)+a_n(t)T}{\Delta x (t)}]^m \frac{\Delta v(t)}{\Delta x^{l-m} (t)}.
\end{equation}
Therefore, we obtain the effect of all leading vehicles on car-following acceleration in car-following model.
In real life, the speed of the driver depends on a closet leading vehicle, that is, the most powerful leading vehicle. In other words, after the projection, it is the car nearest to car-following in the oval . So, we get
%
$
\Delta x(t)=min\{ [(x_n - x'_{n-1})^2+(y_n - y'_{n-1})^2]^{\frac{1}{2}},\cdot \cdot \cdot,[(x_n-x'_{1})^2+(y_n - y'_{1})^2]^{\frac{1}{2}}  \}
$and the corresponding speed difference, where $(x'_{n-i},y'_{n-i}),\ i=1,2,\cdot \cdot \cdot,n-1$ is the leading vehicle coordinate after ellipse projection.


Then, we can get car-following speed in t + T time:


\begin{eqnarray}\label{sudubianhua}     
\left\{                        
\begin{array}{lll}       
u_{n-1}(t+T)=u_{n-1}(t)+ a_{n-1x}(t+T) \\  
v_{n-1}(t+T)=v_{n-1}(t)+ a_{n-1y}(t+T). \\

\end{array}              
\right.                       
\end{eqnarray}

car-following coordinates in $t+T$ time:
\begin{eqnarray}\label{zuobianbianhua}     
\left\{                        
\begin{array}{lll}       
x_{n-1}(t+T)=x_{n-1}(t)+ u_{n-1}(t)T \\  
y_{n-1}(t+T)=y_{n-1}(t)+ v_{n-1}(t)T. \\

\end{array}              
\right.                       
\end{eqnarray}

\subsection{The Choice of Vehicle Direction}

\begin{figure}[!htbp]
\small
\centering
\includegraphics[width=12cm]{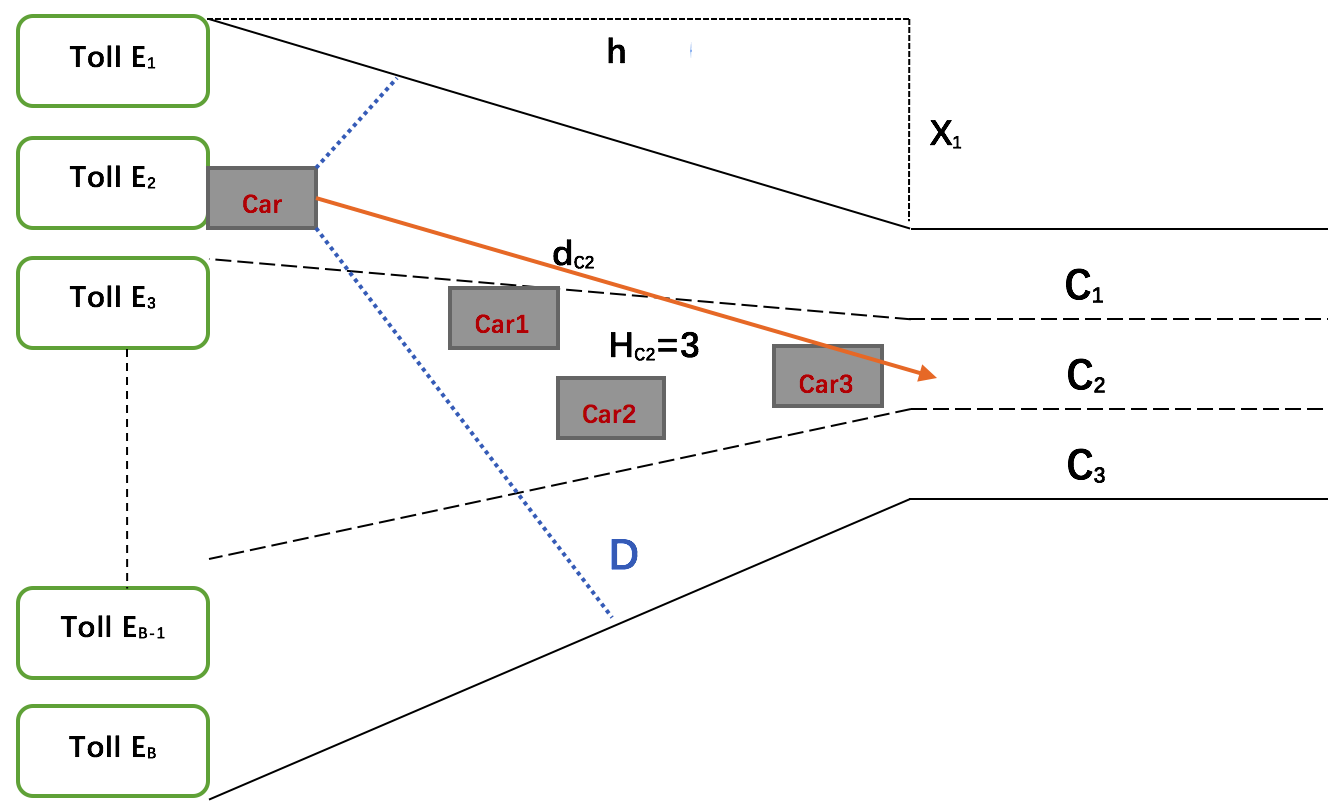}
\caption{:Output system of toll booth(the right side of the two blue dashed lines is the observation area)} \label{shuchushanxing}
\end{figure}

When the vehicle drive out of the toll, into the fan-shaped buffer area, it is about to converge with other vehicles and drive into the trunk highway(see figure \ref{shuchushanxing}). Vehicles need to decide which lane to import, that is, to choose their own direction. First, we define the observed region. $C_i$ is the i-th lane of the expressway, where $i=1,\cdot,\cdot,\cdot,L$. The observation area is a section of the lane $C_i$. At this point the driver will determine the number of queued vehicles in the area. The number of queuing vehicles in the observation area of lane $C_i$ is $H_{Ci}$. For example, there are $H_{C2}=3$ queuing vehicles in the $C_2$ observation area. These vehicles are about to enter the $C_i$ lane. The right side of the two blue dashed lines is the observation area, denoted by D. We then define the straight-line distance of the current vehicle to the intersection of freeway lane $C_i$ as $d_{Ci}$(see figure\ref{shuchushanxing}). Drivers will choose which highway lane to enter. The choice of car will be affected by  $H_{Ci}$ and $d_{Ci}$ .The higher the $H_{Ci}$ is, the higher the number of waiting vehicles in the $C_i$  lane in the observation area is, the lower the probability that the vehicle will choose. The higher the distance $d_{Ci}$ is, the higher the distance from the vehicle to the $C_i$ lane will be, the lower the probability that the vehicle will select the lane will be. The probability of vehicle choosing lane is affected by $H_{Ci}$ and $d_{Ci}$.

\par

We determine the weights of probability of $k_1$ and $k_2$ that $H_{Ci}$ and $d_{Ci}$ affect the vehicle-selected lane through the data. Let $P_i$ be the probability of choosing the i-th lane for the vehicle, namely
\begin{eqnarray}\label{1}
P_i=\frac{1}{k_1 \cdot H_{Ci} +k_2 \cdot d_{Ci}}.
\end{eqnarray}
And the vehicle can select the lane with the highest probability as its forward direction.


\section{Numerical Simulation}

\subsection{Coordinate system}

General Definitions

The width of a toll: $d_t$.

The width of a lane£º$d_l$.

The toll booth number of the exit vehicle at the next time: n.

\begin{figure}[!htbp]
\small
\centering
\includegraphics[width=12cm]{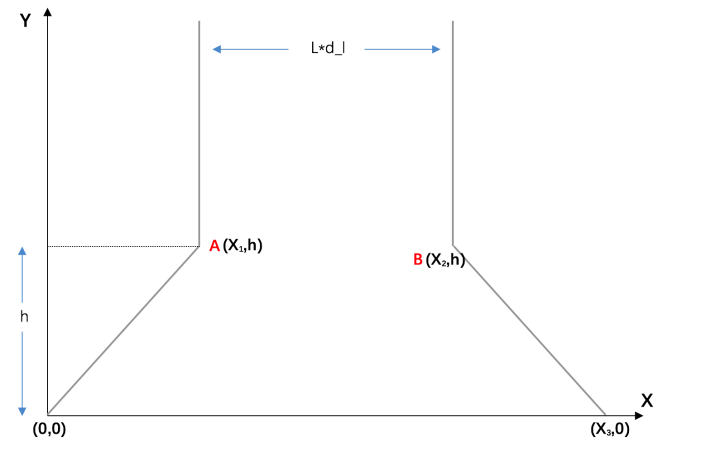}
\caption{:coordinate system of simulation} \label{fangzhenzuobiao}
\end{figure}

To construct a two-dimensional coordinate axis oriented in X and Y directions. The X-axis in the first quadrant represents all the exits of the tollbooth. As shown in the figure\ref{fangzhenzuobiao}, what above points A and B are the confluence of the road. The coordinates of point A and point B are $(X_1,h)$ and $(X_2,h)$, respectively. The total width of the road after merging is $L\cdot d_l$. In the next time, the car coordinates $N(X_n,0)$ which drives out.
From the figure we know abscissa expression of pointsrequired by the simulation is:


\begin{eqnarray} \label{7}    
\left\{                        
\begin{array}{lllll}       
X_n=(n-1)\cdot(d_l+d_t)+\frac{d_l}{2} \\  
X_3= B\cdot(d_l+d_t)\\
X_2=\frac{X_3+L\cdot d_l}{2}\\
X_1=\frac{X_3-L\cdot d_l}{2}.

\end{array}              
\right.                       
\end{eqnarray}

\par

\subsection{Vehicle Arrives System}

In order to analyze the characteristics of traffic flow, it is necessary to simulate the behavior of the arrival of multiple vehicles, and on this basis, we need to make statistics on the average delay time of multiple vehicles through the toll station.

When the arrival of the vehicle satisfies the condition of non-effectiveness, smoothness and universality, the arrival of the vehicle obeys Poisson distribution. In general, the probability of arrival of n vehicles in the period $[t_0,\ t_0+t ]$ satisfies the previously mentioned Poisson distribution formula.



\subsection{Simulation of Vehicle Movement Rules}

We believe that the two factors affecting the movement of vehicles are the vehicle speed and vehicle direction. Therefore, the moving rule of the vehicle is divided into a rule for determining the rate (Rule 1) and a rule for determining the traveling direction (Rule 2).

Rule 1:
As the vehicle is in the psychological field, it will be subject to a number of vehicles around the force. However, considering the real life, the driver's driving speed is more affected by the impact of the nearest vehicle. Therefore, in this simulation, when we update the speed of each vehicle, we only consider the speed of the vehicle closest to its straight line. By (\ref{anxinlichangjianhua}) and (\ref{sudubianhua}), the magnitude of the velocity at the next moment of the vehicle can be determined from the forces of pace car and the quadruple that describe the vehicle state. The specific changes in speed and the method of calculation have been given above.


Rule 2:
At the time of vehicle path calculation, the current running direction of the vehicle is very important. Therefore, it is necessary to consider the influence of all the vehicles in the field during the determination of the forward direction of the vehicle. In determining the process, the impact of each vehicle needs to be superimposed. We find a more simple way is to use the equipotential surface knowledge, project the field of all vehicles¡¯ impact in one direction. The direction of motion of the vehicle at a later time can then be determined by simple scalar computation and orthogonal decomposition.

\begin{figure}[ht]
 \begin{minipage}[ht]{0.5\linewidth}
 \centering
 \includegraphics[width=0.9\textwidth]{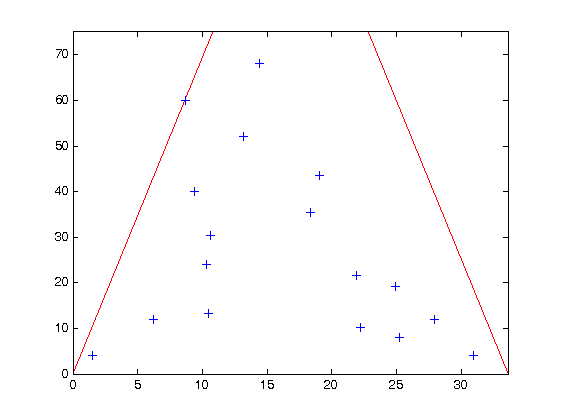}
\caption{:simulation of cars based on low density} \label{low density}
 \end{minipage}
 \begin{minipage}[ht]{0.5\linewidth}
 \centering
 \includegraphics[width=0.9\textwidth]{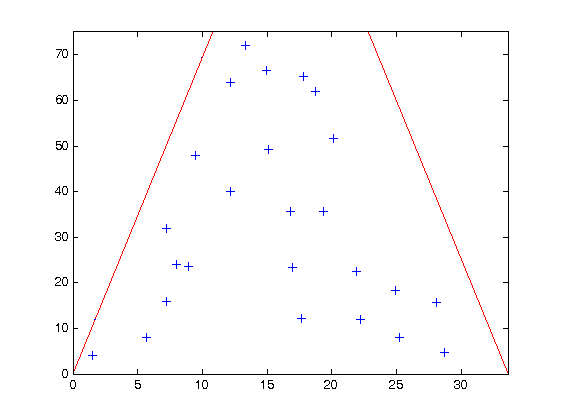}
\caption{:simulation of cars based on high density} \label{high density}
\end{minipage}
\end{figure}

The simulation results show the vehicle movement( see figure\ref{low density} and \ref{high density}). From figures we can know that, when car exits the toll, it will go toward the highway. Meanwhile, it is under the effect of other cars and the boundary of line. Most of cars go toward highway, but there are still some cars deviate from the highway because they are effected by obstacle.

\section{Evaluation Model}

\subsection{Throughput Analysis } 

\begin{figure}[ht]
 \begin{minipage}[ht]{0.5\linewidth}
 \centering
 \includegraphics[width=1\textwidth]{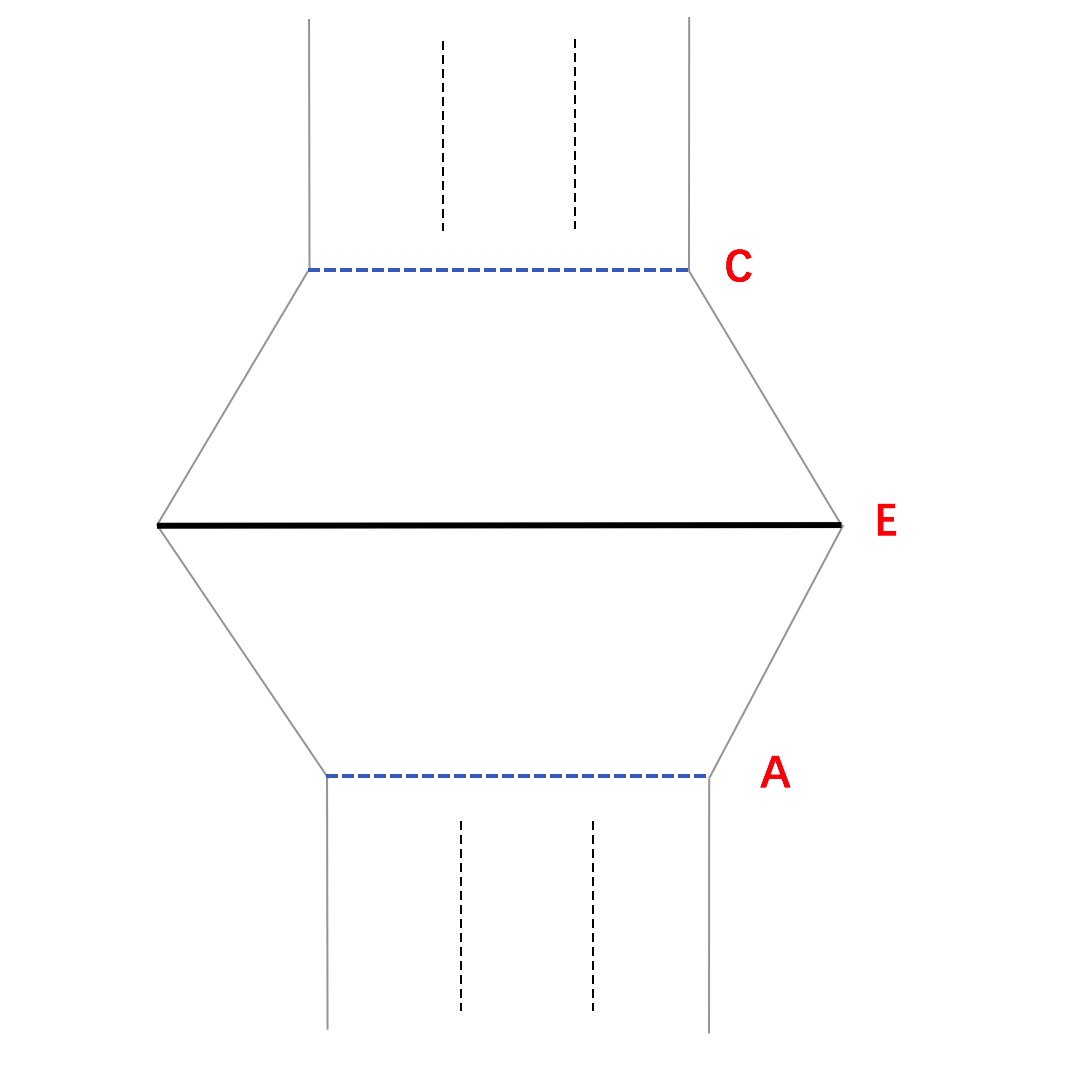}
\caption{:input and output of the toll} \label{tuntuliang}
 \end{minipage}
 \begin{minipage}[ht]{0.5\linewidth}
 \centering
 \includegraphics[width=1.1\textwidth]{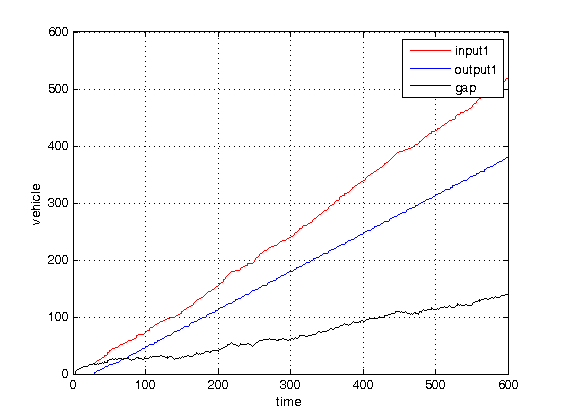}
\caption{:comparation of throughput between A and E} \label{ab}
\end{minipage}
\end{figure}

In evaluating the toll plaza's throughput, the toll plaza can be divided into several sections in the figure below(see figure\ref{tuntuliang}).
The black line E in the figure represents all the tollbooths. The blue dashed line represents the border of the toll plaza exit. When the vehicle exceeds the line segment A, the vehicle enters the toll plaza. When the vehicle exceeds the line segment E, the vehicle passes the toll booth. When the vehicle exceeds the line segment C, the vehicle leaves the toll plaza.


In the simulation process, we can determine the location of the vehicle in the toll plaza through the vertical coordinates of each vehicle. By describing the four-tuple of vehicle states, we can get the number of vehicles passing through these three segments at each moment. And then the distribution function of the vehicle on these three line segments can be obtained. To analyze the number of vehicles per second through the three segments in a period of time, we can determine the toll plaza throughput. The number of cars passing through A per unit time represents the number of cars entering the toll plaza. The number of vehicles passing through E represents the number of cars that have been paid. The number of cars passing through C per unit time represents the number of vehicles leaving the toll plaza.
Through the Matlab simulation process, we get the distribution function image(see figure\ref{ab}).


\begin{figure}[ht]
 \begin{minipage}[ht]{0.5\linewidth}
 \centering
 \includegraphics[width=1.1\textwidth]{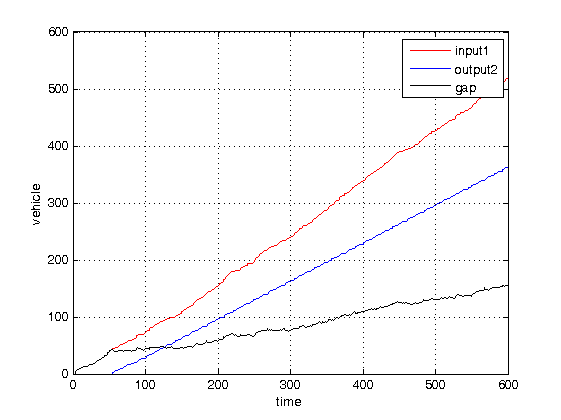}
\caption{:comparation of throughput between A and C} \label{ac}
 \end{minipage}
 \begin{minipage}[ht]{0.5\linewidth}
 \centering
 \includegraphics[width=1.1\textwidth]{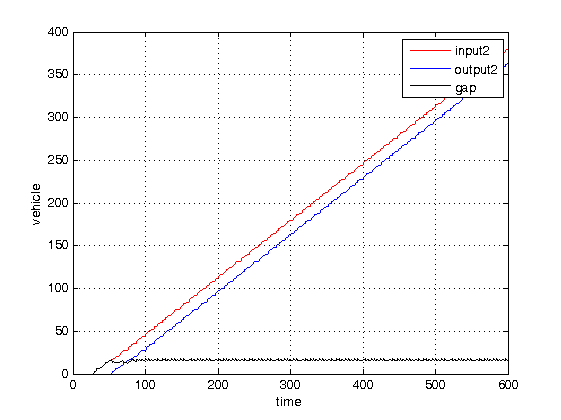}
\caption{:comparation of throughput between E and C} \label{bc}
\end{minipage}
\end{figure}

The figure is an image of the throughput of the portion between lines A and E. The abscissa represents time, and the ordinate represents the number of cars. The red line represents the vehicle distribution function on line A. The blue line represents the vehicle distribution function on line E. For example, the point (100,100) represents the number of cars passing through this line in the first 100 seconds.
The black line is the difference between the distribution functions on the two lines, and the black line can be used to characterize the number of delayed vehicles. The ordinate at each point on the line represents the number of cars remaining between A and E at this point.

As can be seen from the figure, the slope of the blue line is gradually smaller than the slope of the red line, at the same time, the value of the black line gradually increased. This means that the number of cars between segments A and E has been increasing. This shows that between A and E there are queuing vehicles, the situation will increase with time traffic jams.
As same, we can analyze the throughput between E and C situation. That is to analyze the vehicle confluence stage. Through Matlab simulation we get the following figure(see figure\ref{bc}):

It can be seen that in this case the input and output maintain a rather balanced relationship. This image represents the type of toll plaza that we want.

Similarly, you can analyze A, C throughput between the situation, that is, to analyze the toll plaza to make the following throughout analysis.

As can be seen from the figure\ref{ac}, the vehicle input and output are not balanced. As time increases,there are more and more queuing vehicles in the toll plaza. Obviously, the design of the toll plaza is not good enough.

\subsection{Cost Analysis}

Here cost is divided into customer waiting cost, toll booth construction fees and  toll square construction cost.


Due to the adverse effects caused by low efficiency of toll station service, such as fuel consumption and environmental pollution, which are called customer waiting cost, denoted by $C_d$. The waiting cost of the customer is related to the delay time of the vehicle through the toll plaza. Assuming that the average speed of the current highway is v, the total length of the toll plaza is W. The length of the toll plaza includes the length of the fan-over area and the toll station area. Assuming there are no toll station, the normal passage of the road segment time is  $t_1=w/v$. When there is a toll station, the average time the vehicle go through the toll plaza is $t_2$. So we can get the time delay $t_d=t_2-t_1$. Time delay will not only affect the driver, so drivers spend more time waiting for the increase in costs. And it will consume more fuel that generate more energy costs. At the same time, air pollution will become more serious, thus increasing environmental costs. These can be counted in the customer waiting cost, that is, the longer the time delay, the higher the waiting cost is. Hence, we have

\begin{equation}\label{}
C_d=f(t_d)=a_n\ t_d^n+a_{n-1}\ t_d^{n-1},\cdot \cdot \cdot,a_{1}\ t_d+a_0.
\end{equation}

The increase in the number of toll booths will increase the toll station system construction costs and operating costs and operating costs, the cost to build toll booths is $C_t$. Automated Toll Collection (ATC), Manual Toll Collection (MTC), and Electronic Toll Collection (ETC) are not all the same, so the cost of these three types of toll booths varies. The number of ATC is $m_{ATC}$, the number of MTC is $m_{MTC}$, the number of ETC is $m_{ETC}$. At the same time, we note the costs of building a single toll booth are $c_{ATC}$, $c_{MTC}$ and $c_{ETC}$. Therefore, we get

\begin{equation}\label{}
C_t=m_{ATC}\times c_{ATC}+m_{MTC}\times c_{MTC}+m_{ETC}\times c_{ETC}.
\end{equation}

The cost of the toll plaza $(C_s)$  is mainly the square area $(Z)$ $\times$ construction cost per unit area $(c_u)$+ fixed cost $(c_g)$. Square area is mainly the sum of confluence area $Z_1$, diversion zone $Z_2$ and toll station $Z_t$ area. Namely,

\begin{equation}\label{}
C_s=Z\times c_u +c_g=(Z_1+Z_2+Z_t)\times c_u +c_g
\end{equation}
We consider the toll station width, assuming there are B toll channels, the toll station width is(see figure \ref{fangzhenzuobiao}):

\begin{equation}\label{}
X_3=(B+1)d_t+Bd_l+2e,
\end{equation}
where $d_l$ is the toll lane width (m), $d_t$ is the toll booth width (m), and e is the lateral shoulder width and the lateral lane width (m) of the toll plaza.
Toll lane width of toll is 3.0-3.5m, toll booths take the width of 1.8-3.2m, lateral soil shoulder width and lateral lane width take 8.5-9m, toll booth length take more 18-24m\cite{Qiao}. Assume that the confluence highway has L lanes in one direction, each lane width is $d_l'$, and it is 3.75m. So the width of the highway is: $X_2-X_1=d_l'\ L$.


The length of the toll station is $K$. The length of the toll booth depends on the length of the toll booth and the length of the straight line before and after the toll booth. In order to facilitate the vehicle into the toll station after deceleration, parking and start, before and after the toll booth straight line length should be greater than 20m. So the length of the toll station should be greater than 34m. Therefore, we can get the toll station area:

$
Z_t=X_3\times K.
$

Taking into account the fan-shaped transition section design, in order to let the vehicle successfully pass through the toll station, you need to set toll stations and standard high-speed section between the set transition section (fan-shaped gradient area). According to the Japanese highway design methods (1987), the provisions of the gradient rate $X_1/h$ should be less than 1/3\cite{5}(see figure \ref{fangzhenzuobiao}). The longer the length of the toll station transition area is, the smaller the gradient is, the more naturally the vehicle drive from the toll station into the standard high-speed road. We begin to calculate the sector buffer area. We can regard the upper boundary of the sector buffer as a piecewise function,

\begin{eqnarray} \label{7}    
f(x)=\left\{                        
\begin{array}{lllll}       
f_1(x),x\in[0,X_1] \\  
h,\ x\in[X_1,X_2]\\
f_2(x),x\in[X_2,X_3] .

\end{array}              
\right.                       
\end{eqnarray}
Therefore we get the fan-shaped area,

\begin{eqnarray}
Z_2=\int_0^{X_3}f(x)dx.
\end{eqnarray}Similarly, we can obtain $Z_1$.

From this we can obtain customer waiting cost, toll booth construction cost and toll plaza construction cost. How to consider these three aspects of the cost, making the sum of the three types of costs to a minimum, is what the toll station settings should be considered.


\section{Results Analysis}

Analog simulation is realized by means of MATLAB. The number of lanes is denoted by L, the number of tollbooths is denoted by B, and the vertical distance between the exit of toll plaza and tollbooths (see figure\ref{fangzhenzuobiao}) is denoted by h. These variables affect the shape and size of the cushion area. After simulation, the throughput, cost and safety factor are obtained. The results are shown in table\ref{simulation}.


\begin{table}[!htb]
\centering
\caption{:The result of simulation }\label{simulation}
\begin{center}
\begin{tabular}{ccc|ccc}
\hline
L & B & h & throughput & cost &  safety factor \\
\hline

2& 6& 50	& 4320	& 603000& 	0.0154\\
2& 6& 75	& 4980	& 856500& 	0.0145\\
2& 6& 100	& 4260	& 1110000& 	0.0179\\
2& 7& 50	& 4500	& 687250& 	0.0169\\
2& 7& 75	& 4740	& 974880& 	0.0166\\
2& 7& 100	& 4500	& 1262500& 	0.0158\\
2& 8& 50	& 4560	& 771500& 	0.0131\\
2& 8& 75	& 4620	& 1093250& 	0.0168\\
2&8&100	& 4560	& 1415000& 	0.0131\\
3&6&50	& 5400	& 651750& 	0.0211\\
3&6&75	& 5460	& 929625& 	0.0203\\
3&6&100	& 5400	& 1207500& 	0.0216\\
3 & 7      & 75      &6300  & 1048000 &  0.0215      \\
3& 7& 75	& 6300	& 1048000	& 0.0215\\
3& 7& 50	& 6300& 	736000& 	0.0198\\
3& 7& 100	& 6360	& 1360000	& 0.0219\\
3&8&50	& 6900	& 820250& 	0.0235\\
3&8&75	& 6240	& 1166375& 	0.0207\\
3&8&100	& 6420	& 1512500& 	0.0211\\
4&6&50	& 5340	& 700500& 	0.0194\\
4&6&75	& 5460	& 1002750& 	0.0186\\
4&6&100	& 5400	& 1305000& 	0.0199\\
4&7&50	& 6360	& 784750& 	0.0193\\
4&7&75	& 6240	& 1121125& 	0.0206\\
4&7&100	& 6180	& 1457500& 	0.0213\\
4&8&50	& 7320	& 869000& 	0.0211\\
4&8&75	& 7200	& 1239500& 	0.0211\\
4&8&100	& 7260	& 1610000& 	0.0219\\
 \hline
\end{tabular}
\end{center}
\end{table}

The establishment of evaluation model is as follows. In order to reflect the capacity pressure under different traffic demand, we construct the following formula

\begin{equation}\label{}
FFF=min\{X- F(B,\ L,\ S),0\},
\end{equation}

where X is the maximum traffic demand, F(B, L, h) is the maximum throughput under various B, L and h. When X is greater than F(B, L, h), the capacity pressure can be represented by the difference between them; when X is less than F(B, L, h), the capacity pressure is 0, which means there is no capacity pressure. Before comprehensive evaluation, the data set to be evaluated need to be processed into dimensionless data:
\begin{equation}\label{wuliangganghua}
y_i=\frac{x_i-x_min}{x_max-x_min},
\end{equation}
where $x_{min}$ is the minimum sample, $x_{max}$ is the maximum sample. The evaluation function is:
\begin{equation}\label{pingjiahanshu}
Y=A_1X_1+A_2X_2+A_3(X_3),
\end{equation}
where $X_1$  is capacity pressure,  $X_2$ is construction cost, $X_3$ is safety factor. The proportional coefficients of them are 0.4, 0.2 and 0.4. After processing by MATLAB, we obtain the optimal solutions under various B, L and h (see table\ref{butongcheliuliang}).


\begin{table}[!htb]
\centering
\caption{:The simulation results of different range of maximum traffic flow }\label{butongcheliuliang}
\begin{center}
\begin{tabular}{c|ccc}
\hline
The range of maximum traffic flow & L & B & h \\%
\hline
$4000-4600$ &2 & 8 & 50  \\
$4700-5500$ & 2 & 6 & 75  \\
$5600-6800$ & 4 & 7 & 50  \\
$6900-7800$ & 4 & 8 & 50  \\
\hline
\end{tabular}
\end{center}
\end{table}

We define safety factor as the number of vehicles per square meter. For security reasons, the proportional coefficient of safety factor is greater, which means the optimal solutions must ensure vehicle security. From table 2, we can tell that the number of tollbooths is a key factor in light traffic, which means optimal solutions have more tollbooths. Consequently, the optimal solution has 8 tollbooths. However, with the increasing traffic flow, congestion may occur at the exit of cushion area and more traffic accident may appear if the number of lanes does not change. Therefore, the number of lanes needs to be increased. As shown in table\ref{butongcheliuliang}, the number of lanes of optimal solution has changed from 2 to 4.

In the following, when the range of traffic flow is given, we choose the best model and change the proportion of human-staffed tollbooths(MTC), automated tollbooths(ATC), electronic toll collection booths(ETC).

We consider the situation when the maximum traffic flow ranges from 6900 ¨C 7800. The optimal solution under this situation can be represented by $(L,B,h)=(4,8,50)$. We change the proportion of MTC, ATC and ETC and obtain results shown in table\ref{butongbilitoll}).

\begin{table}[!htb]
\centering
\caption{: The simulation results: different proportion of 3 kinds of tollbooths}\label{butongbilitoll}
\begin{tabular}{cccccc}
\hline
MTC & ATC & ETC & throughput & cost &  safety factor \\
\hline
8&0&0	&2400	&869000	&0.0064\\
0&8&0	&3900	&997000	&0.0135\\
0&0&8	&8100	&1509000	&0.0234\\
4&4&0	&3180	&933000	&0.0105\\
4&0&4	&8400	&1189000	&0.0257\\
0&4&4	&8580	&1253000	&0.0263\\
2&3&3	&7380	&1157000	&0.0246\\

\hline
\end{tabular}
\end{table}
By analyzing the data in table\ref{butongbilitoll}, we find that ETC and ATC has a shorter service time length than MTC, so throughput can be increased by adding the proportion of ETC/ATC properly. However, limited by the number of lanes, vehicles may form into congestion at the exit of cushion area when traffic flow increases, which cannot be fixed by adding the proportion of ETC/ATC. Such congestion not only restricts throughput but also leads to insecurity.


\section{Sensitivity Analysis Based on Autonomous Vehicles}
\begin{figure}[!htbp]
\small
\centering
\includegraphics[width=8cm]{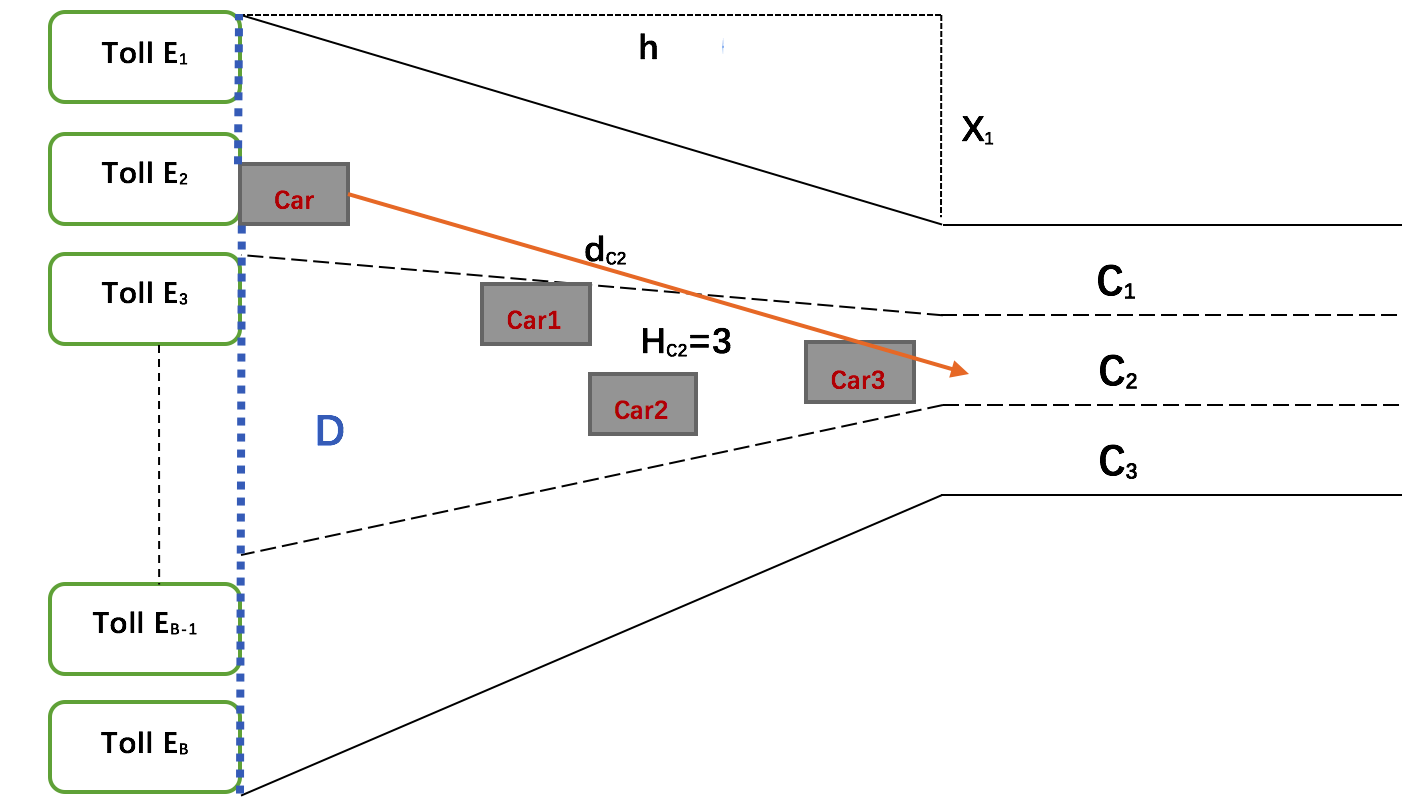}
\caption{:The observation area of autonomous vehicles which is right side of blue dashed line} \label{wurenche}
\end{figure}

We need to test the sensitivity of our model. We let more autonomous vehicles get into our system. Note that autonomous vehicle is controlled by computer system and the system can consider all autonomous vehicles at the same time. Thus, we need to make some changes to our model, namely, the observation area need to change(the conventional observation area is based on driver,see figure\ref{shuchushanxing}). Now, vehicles are under the control of computer and it can observe the whole area(see figure\ref{wurenche} which is right side of blue dashed line). Hence, the observation area autonomous vehicle is the whole system. The observation area determine the direction of vehicle. Similarly, interaction between vehicles need to change. The conventional interaction between vehicles just consider the largest force that effect the vehicle(the detail analysis is in section5.2). Now, we need to consider all the force that effect the vehicle. The interaction between vehicles determine the accelerated velocity. Therefore, we apply these new to our model.
\par

\section{Conclusion}

We build two-dimensional traffic model based on psychological field. According to the force analysis, we get the speed rule and heading rule of cars. We use Matlab to do numerical simulation. We get the optimal solution of different range of traffic flow, namely, the optimal value of throughput, cost and accident factor based on evaluation model. When the traffic flow belongs to 4000-4600, it is better to have 2 lines of a toll highway, 8 tollbooths and the height of fan is 50m. When traffic flow goes higher, we need to increase lines of a toll highway.

Further, we change the proportion of different kinds of tolls, we have, if we increase the proportion of ETC appropriately, we can let throughput goes higher. But it is still under the restrictions of highway lines.


\begin{thebibliography}{30}
\bibitem{Campbell} Campbell P J. Author¡¯s Commentary: The Outstanding Exhaustible Resource Papers[J]. UMAPJournal, 2005: 179.
\bibitem{Wilson}Wilson, R. A. (2000). Transportation in America: 1999. Statistical Analysis of Transportation in the United States: 17th.
\bibitem{Rillings} Rillings J H. Automated highways[J]. Scientific American, 1997, 277(4): 80-85.
\bibitem{Rong} Rong J. The capacity of expressway toll-gate on the basis of M/G/K queuing model in Beijing Region[J]. Highway, 2001 (7): 128-133.
\bibitem{Prigogine} Prigogine I, Herman R. Kinetic theory of vehicular traffic[R]. 1971.
\bibitem{Lighthill} Lighthill M J, Whitham G B. On kinematic waves. I. Flood movement in long rivers[C]//Proceedings of the Royal Society of London A: Mathematical, Physical and Engineering Sciences. The Royal Society, 1955, 229(1178): 281-316. MLA 
\bibitem{Whitham}  Lighthill M J, Whitham G B. On kinematic waves. II. A theory of traffic flow on long crowded roads[C]//Proceedings of the Royal Society of London A: Mathematical, Physical and Engineering Sciences. The Royal Society, 1955, 229(1178): 317-345.
\bibitem{YAO}  YAO Ronghan£® A Study on Vehicular Queue Models £ÛD£Ý£® Changchun: Jilin University£¬2007.
\bibitem{Garber}  Garber N J, Hoel L A. Traffic and highway engineering[M]. Cengage Learning, 2014.
\bibitem{Nagel K}  Nagel K, Schreckenberg M. A cellular automaton model for freeway traffic[J]. Journal de physique I, 1992, 2(12): 2221-2229.














\bibitem{Edie} Edie A C. Traffic delays at toll booths [J]. Journal of Operations Research Society of America. 1954, 2:107-138.
\bibitem{Houston}Texas Transporation Institute. Houston's travel rate improvement prgram [R]. http:// mobility. tamu. edu/ums/trip/toolbox/increase system efficiency. pdf,2001.
\bibitem{Tomer} Tomer  Toldeo,  Haris  N.  Koutsopoulos,  Moshe  Ben-Akiva.  Estimation  of  an  integrated  driving
behavior model [J]. transportation Research Part C, 2009, 17(4): 365-380.
\bibitem{Mark} Mark Brackstone, Mike Mcdonald. Car-following: a historical review[J]. Transprotation Research
Part F, 1999, 2(4)£º181-196.
\bibitem{E Kometani} E Kometani, T Sasaki. DYNAMIC BEHAVIOUR OF TRAFFIC WITH A NON-LINEAR SPACING- SPEED RELATIONSHIP[J]. Theory of TRAFFIC FLOW, PROCEEDINGS. 1900, 6 : 105-119.
\bibitem{Ajay K} Ajay K. Rathi, Member, ASCE and Alberto J. Santiago. Urban Network Traffic Simulations: TRAF-NETSIM Program[J]. Journal of Transportation Engineering.1990, 116(6) : 734-743.


\bibitem{Tao}  Tao Pengfei.  Modeling of driving behavior based on the psychology field theory [D].Changchun, Jilin University, 2012.

\bibitem{Traffic safety psychology} Wang Jian. Traffic safety psychology [M]. Science and technology literature press .1989.

\bibitem{Qiao}Qiao Xiang.Highway overpass planning and design practice[M].China Communications Press. 2002.
\bibitem{5} Su Li, Study on the design of highway toll station[D].Xi¡¯an, China: Chang'an University, 2012
.Acta Mathematicae App licatae Sinica. 2008, 24(2):195-202.




\end{thebibliography}
\end{document}